\documentclass[sigconf,authorversion]{acmart}
\acmSubmissionID{1234}

\usepackage{booktabs} 

\usepackage{graphicx}
\usepackage{booktabs} 
\usepackage{nicefrac}
\usepackage{pifont}
\usepackage{tikz}
\usepackage{pgfplots}
\usetikzlibrary{shadows,backgrounds,arrows,positioning,matrix,calc}
\pgfplotsset{compat=newest}
\tikzset{
    >=stealth',
    pil/.style={
           ->,
           thick,
           shorten <=2pt,
           shorten >=2pt,}
}

\definecolor{myblue}{rgb}{0.95,0.95,1.0}
\definecolor{mybluelight}{rgb}{0.975,0.975,1.0}
\definecolor{myred}{rgb}{0.9,0,0}
\definecolor{mygreen}{rgb}{0,0.7,0}

\usepackage[ruled]{algorithm2e} 
\usepackage{algorithmic}

\SetAlFnt{\small}
\SetAlCapFnt{\small}
\SetAlCapNameFnt{\small}
\SetAlCapHSkip{0pt}

\setcopyright{rightsretained}
\copyrightyear{2021}
\acmYear{2021}
\acmConference{SIGGRAPH '21 Talks}{August 09-13, 2021}{Virtual Event, USA}
\acmBooktitle{Special Interest Group on Computer Graphics and Interactive Techniques Conference Talks (SIGGRAPH '21 Talks), August 09-13, 2021}
\acmDOI{10.1145/3450623.3464685}
\acmISBN{978-1-4503-8373-8/21/08}

\begin{document}
\title{Passing Multi-Channel Material Textures to a 3-Channel Loss}

\author{Thomas Chambon}
\affiliation{%
  \institution{Unity Technologies}
  \country{}
  \city{}
}
\author{Eric Heitz}
\affiliation{%
 \institution{Unity Technologies}
  \country{}
  \city{}
}
\author{Laurent Belcour}
\affiliation{%
  \institution{Unity Technologies}
  \country{}
  \city{}
}
\renewcommand\shortauthors{Chambon et al.}

\begin{teaserfigure}
\centering
\begin{tikzpicture}
\draw[rounded corners=10pt,fill=myblue] (-6,-1.25+0.75) rectangle (-4,-1.25-0.75);
\draw [->] (-4, -1.25) -- (-3.7, -1.25);
\draw (-5.0,-0.80) node {\small generator};
\draw (-5.0,-1.40) node {\includegraphics[width=0.05\linewidth]{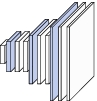}};
\draw[rounded corners=10pt,fill=myblue] (-6,+1.25+0.5) rectangle (-4,+1.25-0.5);
\draw (-5.0,+1.25) node {\small example};
\draw [->] (-4, +1.25) -- (-3.7, +1.25);
\draw[rounded corners=10pt,fill=myblue] (-1,+0.5) rectangle (1,-0.5);
\draw (0,0) node {\small random triplet};
\draw (0,+1.25) circle (0.1);
\draw (0,-1.25) circle (0.1);
\draw [->] (-1.3, +1.25) -- (-0.1, +1.25);
\draw [->] (0, 0.5) -- (0, +1.15);
\draw [->] (0.1, +1.25) -- (1.3, +1.25);
\draw [->] (-1.3, -1.25) -- (-0.1, -1.25);
\draw [->] (0, -0.5) -- (0, -1.15);
\draw [->] (0.1, -1.25) -- (1.3, -1.25);
\draw (-2.5,+1.25) node {\includegraphics[height=0.1\linewidth]{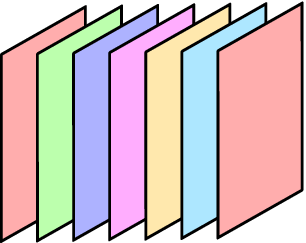}};
\draw (-2.5,-1.25) node {\includegraphics[height=0.1\linewidth]{figures/teaser/input_n_channels.eps}};
\draw (2,+1.25) node {\includegraphics[height=0.1\linewidth]{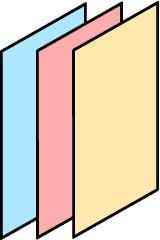}};
\draw (2,-1.25) node {\includegraphics[height=0.1\linewidth]{figures/teaser/input_3_channels.eps}};
\draw (-2.5,0) node {\small $n$ channels};
\draw (2,0) node {\small $3$ channels};
\draw [->] (2.7, +1.25) -- (3, +1.25);
\draw [->] (2.7, -1.25) -- (3, -1.25);
\draw[rounded corners=10pt,fill=myblue] (3,2) rectangle (4.9,-2);
\draw (3.90, 0) node {\includegraphics[width=0.08\linewidth]{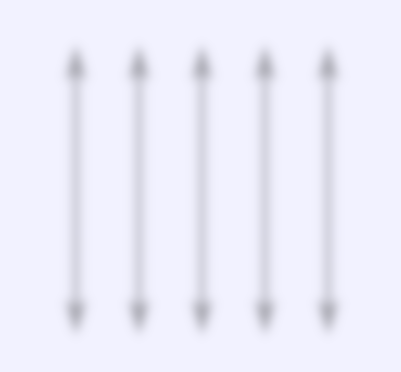}};
\draw (3.90, -1.25) node {\includegraphics[width=0.08\linewidth]{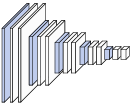}};
\draw (3.90,+1.25) node {\includegraphics[width=0.08\linewidth]{figures/teaser/vgg19.eps}};
\draw (3.95,+0.0) node {\small $3$-channel loss};
\draw (3.95,-0.3) node {\tiny \cite{Gatys15}};
\draw (3.75+0.6, -1.25+0.55) node {\tiny pretrained};
\draw (3.75+0.6, +1.25+0.55) node {\tiny pretrained};
\draw (3.75+0.6, -1.25+0.35) node {\tiny VGG-19};
\draw (3.75+0.6, +1.25+0.35) node {\tiny VGG-19};
\draw [-] (5.10, -2) -- (5.10, +2);
\draw (6.75, +2.05 - 0.2) node {\small example};
\draw (10, 2.05- 0.2) node {\small generated};
\draw (6.83, +0.35- 0.2) node[rectangle,draw,inner sep=0] {\includegraphics[width=0.17\linewidth]{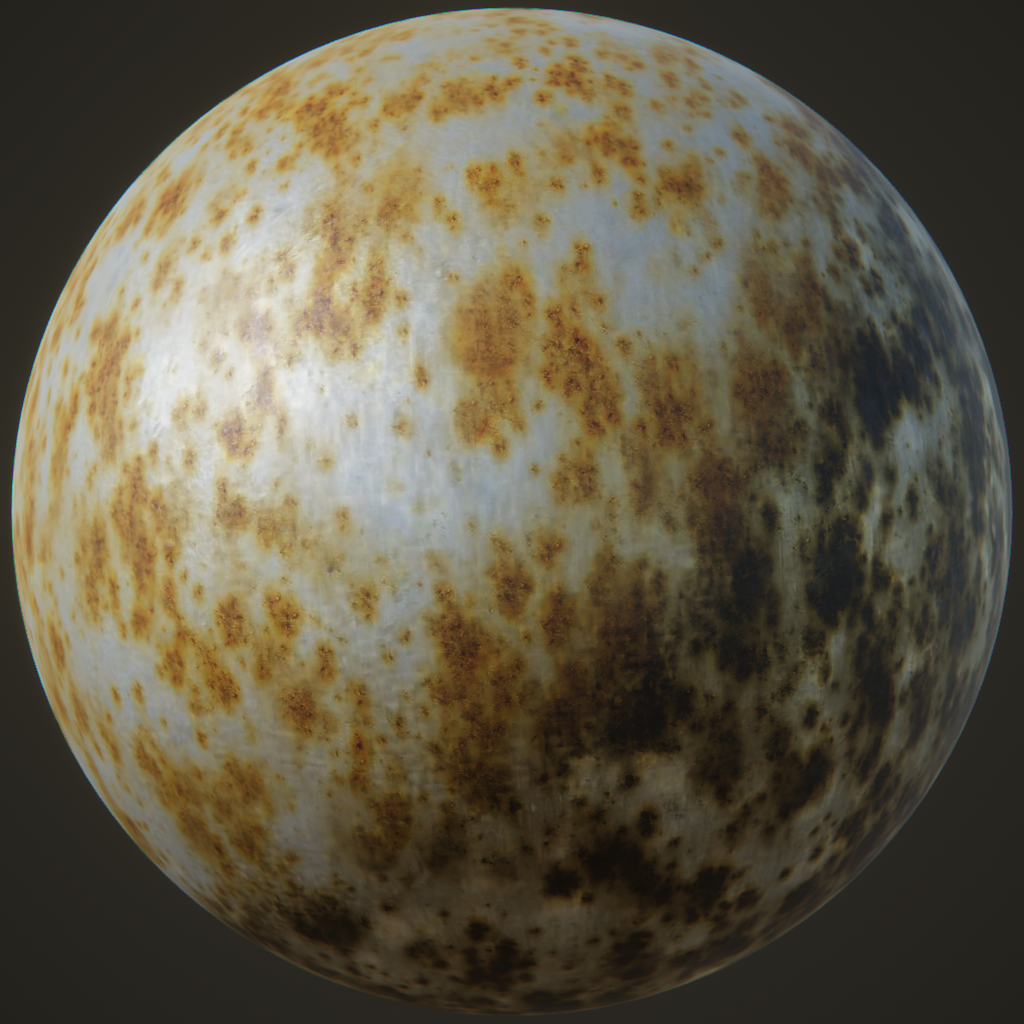}};
\draw (10, +0.35- 0.2) node[rectangle,draw,inner sep=0] {\includegraphics[width=0.17\linewidth]{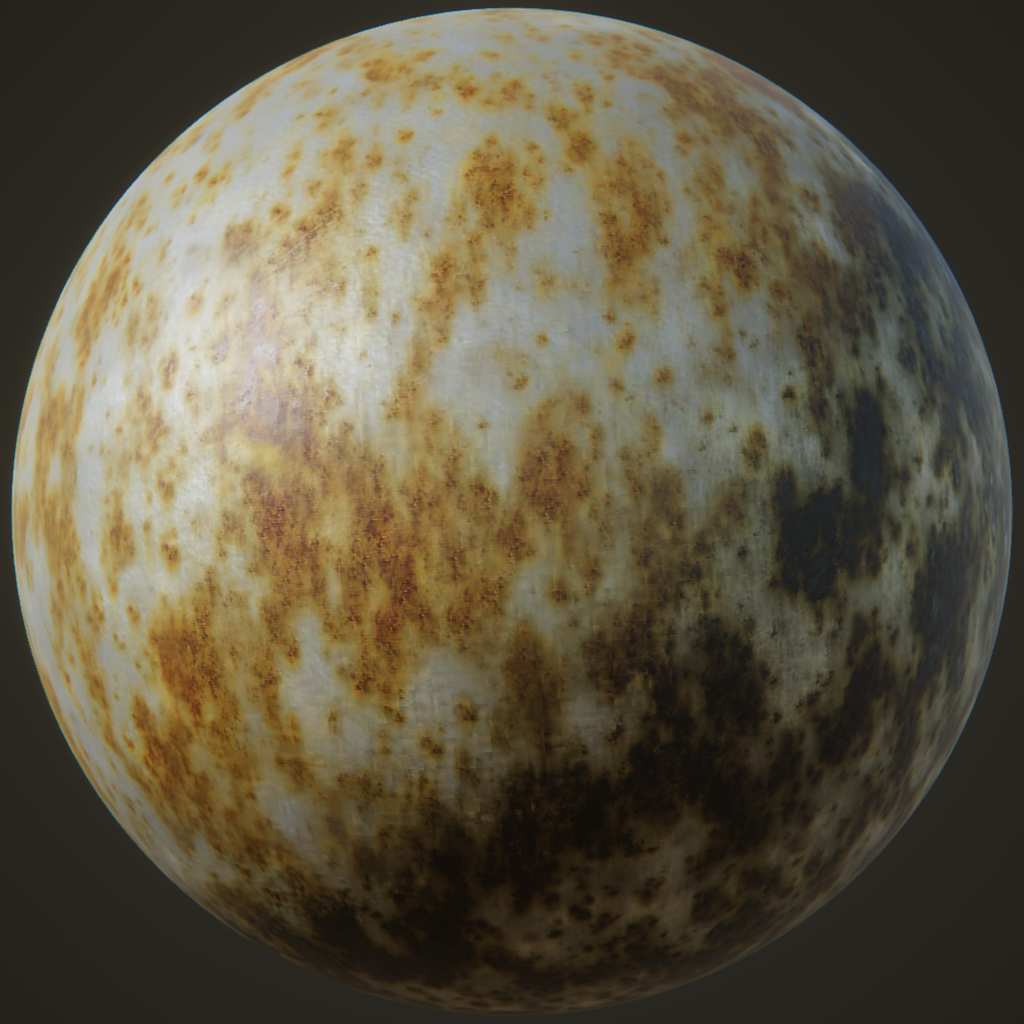}};
\draw (6.83-2*0.62, -1.4- 0.30) node[rectangle,draw,inner sep=0] {\includegraphics[width=0.032\linewidth]{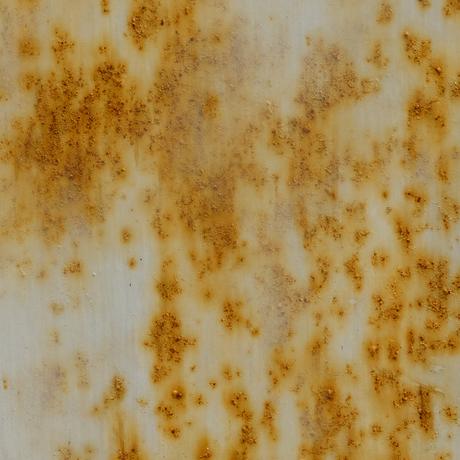}};
\draw (6.83-1*0.62, -1.4- 0.30) node[rectangle,draw,inner sep=0] {\includegraphics[width=0.032\linewidth]{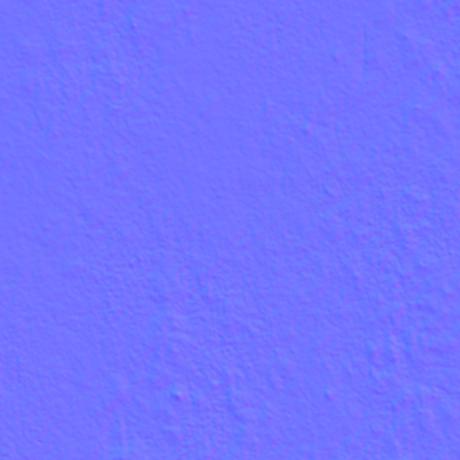}};
\draw (6.83-0.0*0.62, -1.4- 0.30) node[rectangle,draw,inner sep=0] {\includegraphics[width=0.032\linewidth]{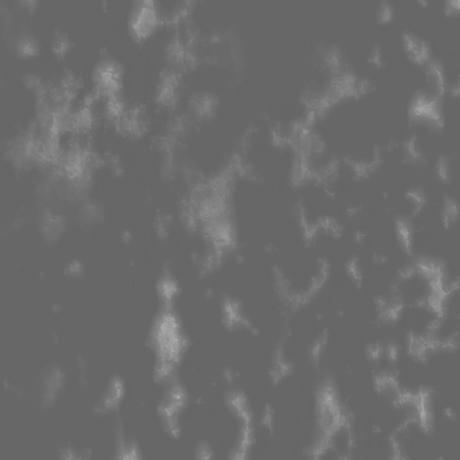}};
\draw (6.83+1*0.62, -1.4- 0.30) node[rectangle,draw,inner sep=0] {\includegraphics[width=0.032\linewidth]{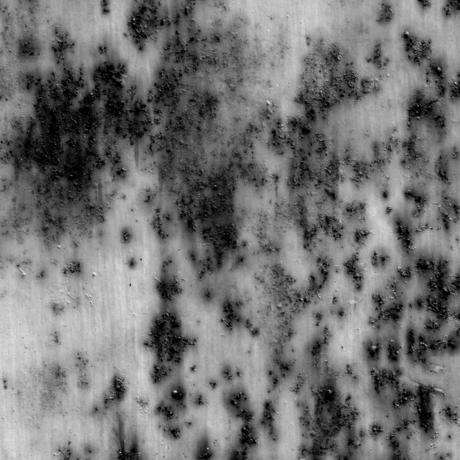}};
\draw (6.83+2*0.62, -1.4- 0.30) node[rectangle,draw,inner sep=0] {\includegraphics[width=0.032\linewidth]{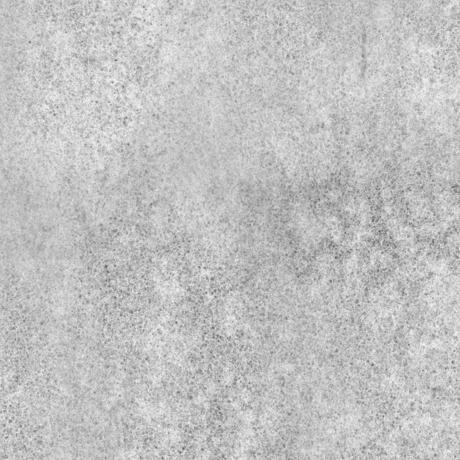}};
\draw (6.83-2*0.62, -1.72- 0.35) node {\scalebox{0.7}{\tiny albedo}};
\draw (6.83-1*0.62, -1.72- 0.35) node {\scalebox{0.7}{\tiny normal}};
\draw (6.83-0*0.62, -1.733- 0.35) node {\scalebox{0.7}{\tiny roughness}};
\draw (6.83+1*0.62, -1.72- 0.35) node {\scalebox{0.7}{\tiny metalness}};
\draw (6.83+2*0.62, -1.72- 0.35) node {\scalebox{0.7}{\tiny ambient}};
\draw (10-2*0.62, -1.4- 0.30) node[rectangle,draw,inner sep=0] {\includegraphics[width=0.032\linewidth]{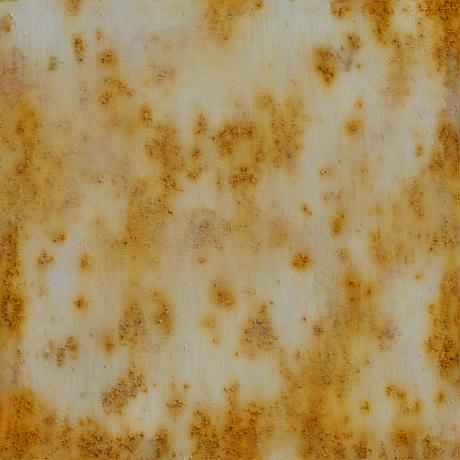}};
\draw (10-1*0.62, -1.4- 0.30) node[rectangle,draw,inner sep=0] {\includegraphics[width=0.032\linewidth]{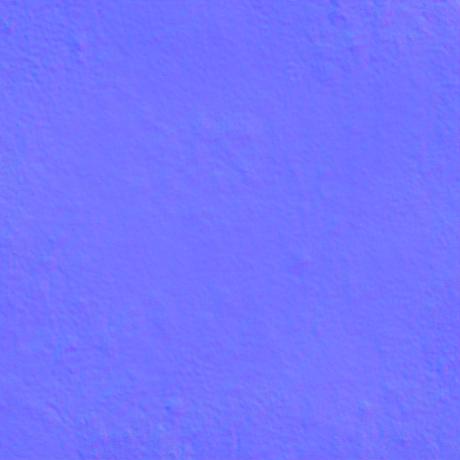}};
\draw (10-0.0*0.62, -1.4- 0.30) node[rectangle,draw,inner sep=0] {\includegraphics[width=0.032\linewidth]{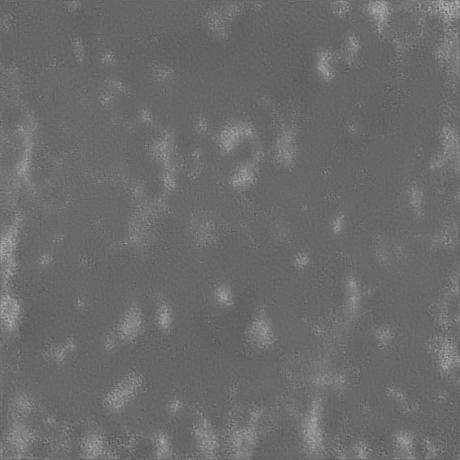}};
\draw (10+1*0.62, -1.4- 0.30) node[rectangle,draw,inner sep=0] {\includegraphics[width=0.032\linewidth]{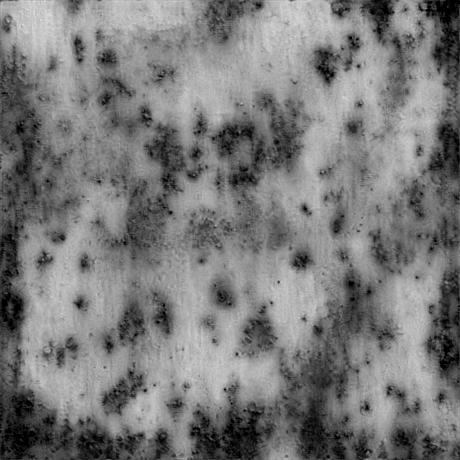}};
\draw (10+2*0.62, -1.4- 0.30) node[rectangle,draw,inner sep=0] {\includegraphics[width=0.032\linewidth]{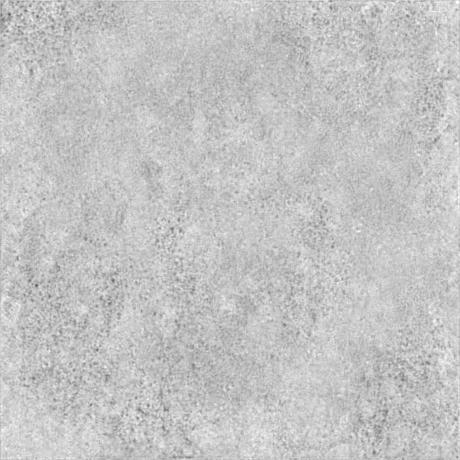}};
\draw (10-2*0.62, -1.72- 0.35) node {\scalebox{0.7}{\tiny albedo}};
\draw (10-1*0.62, -1.72- 0.35) node {\scalebox{0.7}{\tiny normal}};
\draw (10-0*0.62, -1.733- 0.35) node {\scalebox{0.7}{\tiny roughness}};
\draw (10+1*0.62, -1.72- 0.35) node {\scalebox{0.7}{\tiny metalness}};
\draw (10+2*0.62, -1.72- 0.35) node {\scalebox{0.7}{\tiny ambient}};
\end{tikzpicture}
\vspace{-3mm}
\caption{\label{fig:teaser} 
We train a material texture generator with multiple channels such as \emph{albedo}, \emph{normal}, \emph{roughness}, \emph{metalness} and \emph{ambient occlusion} by passing random channel triplets to the $3$-channel loss proposed by Gatys et al.~\shortcite{Gatys15}.
}
\end{teaserfigure}

\begin{abstract}
Our objective is to compute a textural loss that can be used to train texture generators with multiple material channels typically used for physically based rendering such as \emph{albedo}, \emph{normal}, \emph{roughness}, \emph{metalness}, \emph{ambient occlusion}, etc. 
Neural textural losses often build on top of the feature spaces of pretrained convolutional neural networks.
Unfortunately, these pretrained models are only available for 3-channel RGB data and hence limit neural textural losses to this format.
To overcome this limitation, we show that passing random triplets to a 3-channel loss provides a multi-channel loss that can be used to generate high-quality material textures.
\end{abstract}

%
%

%
%


\maketitle

\vspace{-2mm}
\section{Introduction}
\label{sec:introduction}

A neural textural loss allows for generating textures by image optimization~\cite{Gatys15} or training generative models~\cite{UlyanovV1}.
Typically, the loss is computed from the statistics of the feature activations in pretrained Convolutional Neural Networks (CNNs) such as VGG-19 \cite{Simonyan14}. 
These pretrained CNNs are mainly available for RGB inputs, i.e. a $3$-channel formats. 
This is limiting for material textures used in physically based rendering that have multiple channels such as \emph{albedo}, \emph{normal}, \emph{roughness}, \emph{metalness}, \emph{ambient occlusion}, etc.
We thus investigate how a 3-channel loss can be applied to $n$-channel textures.
 
\begin{figure}[h!]
\begin{tabular}{@{\hspace{-1mm}} c @{\hspace{0.35mm}} c @{\hspace{0.35mm}} c @{\hspace{0.35mm}} c @{\hspace{0.75mm}} | @{\hspace{0.75mm}} c @{\hspace{0.35mm}} c @{\hspace{0.35mm}} c @{\hspace{0.35mm}} c @{}}
&
\tiny \textit{albedo} &
\tiny \textit{normal} &
\tiny \textbf{rendered} &
\tiny \textit{albedo} &
\tiny \textit{normal} &
\tiny \textit{roughness} &
\tiny \textbf{rendered} \\
\raisebox{2mm}{\rotatebox{90}{\tiny example}}&
\fbox{\includegraphics[width=0.13\linewidth]{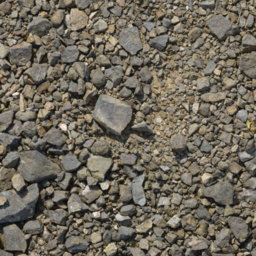}}&
\fbox{\includegraphics[width=0.13\linewidth]{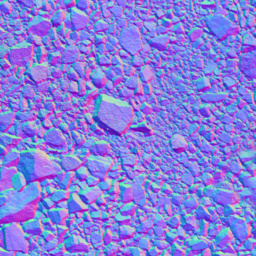}}&
\fbox{\includegraphics[width=0.13\linewidth]{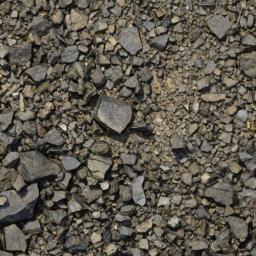}}&
\fbox{\includegraphics[width=0.13\linewidth]{figures/differentiable_renderer/rock_ground_02_diff_2k-0-0-.jpg}}&
\fbox{\includegraphics[width=0.13\linewidth]{figures/differentiable_renderer/rock_ground_02_nor_2k-0-0-.jpg}}&
\fbox{\includegraphics[width=0.13\linewidth]{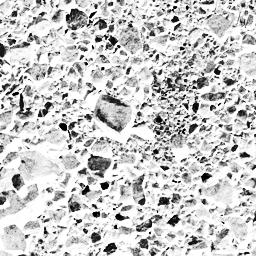}}&
\fbox{\includegraphics[width=0.13\linewidth]{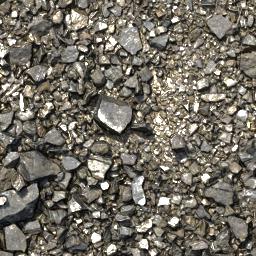}} 
\\
\raisebox{0mm}{\rotatebox{90}{\tiny (a) diff. render}}&
\fbox{\includegraphics[width=0.13\linewidth]{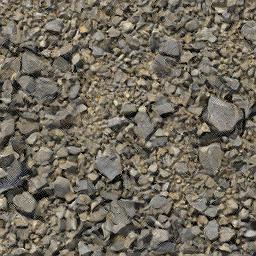}}&
\fbox{\includegraphics[width=0.13\linewidth]{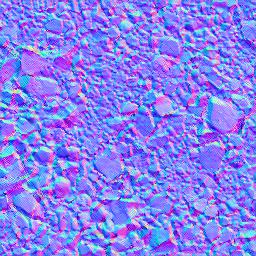}}&
\fbox{\includegraphics[width=0.13\linewidth]{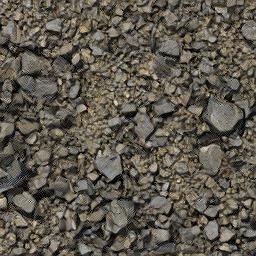}}&
\fbox{\includegraphics[width=0.13\linewidth]{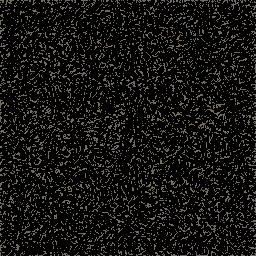}}&
\fbox{\includegraphics[width=0.13\linewidth]{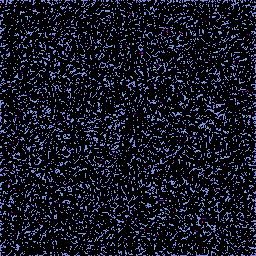}}&
\fbox{\includegraphics[width=0.13\linewidth]{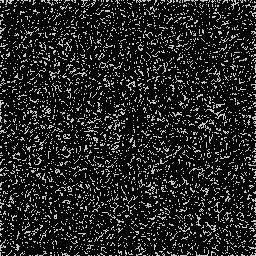}}&
\fbox{\includegraphics[width=0.13\linewidth]{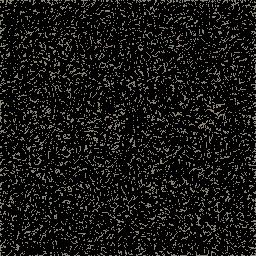}} 
\\
\raisebox{2.5mm}{\rotatebox{90}{\tiny (b) ours}}&
\fbox{\includegraphics[width=0.13\linewidth]{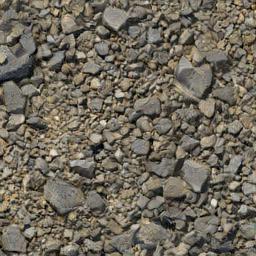}}&
\fbox{\includegraphics[width=0.13\linewidth]{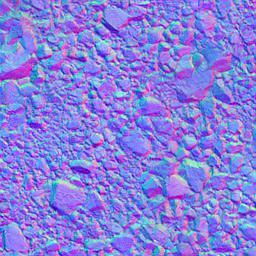}}&
\fbox{\includegraphics[width=0.13\linewidth]{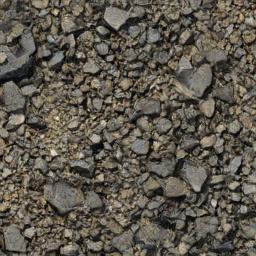}}&
\fbox{\includegraphics[width=0.13\linewidth]{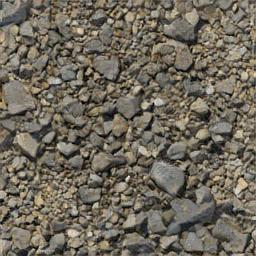}}&
\fbox{\includegraphics[width=0.13\linewidth]{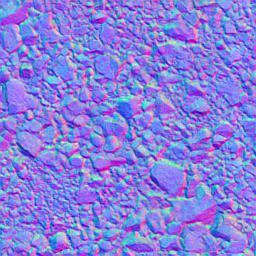}}&
\fbox{\includegraphics[width=0.13\linewidth]{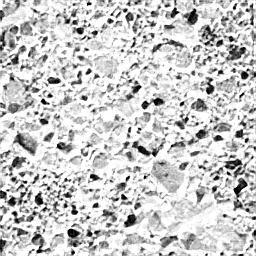}}&
\fbox{\includegraphics[width=0.13\linewidth]{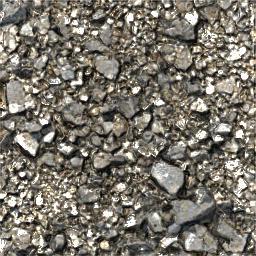}} 
\vspace{-2mm} \\
& \multicolumn{3}{c}{\tiny{diffuse-only material}} & \multicolumn{4}{c}{\tiny{diffuse and rough-specular material}} 
\end{tabular}
\vspace{-5mm}
\caption{\label{fig:differential_rendering_loss} 
(a) Passing a differentiable render to a 3-channel loss works with diffuse materials but becomes unstable with rough speculars. 
(b) Our approach is robust and stable. 
}
\vspace{-10mm}
\end{figure}

Our first attempt was inspired by previous work that generates material textures from RGB photographs examples~\cite{aittala2016}.
They use a differentiable renderer to light the material textures and create an RGB render that can be passed to an RGB loss.
The material textures can then be optimized via gradient descent by backpropagating gradients through the differential renderer.
As shown in Figure~\ref{fig:differential_rendering_loss}-(a), we found this approach to be unstable with non-diffuse materials, especially sharp speculars.
Indeed, in addition to texture synthesis, the optimizer also needs to solve a challenging inverse rendering problem. 
Aittala et al.~\shortcite{aittala2016} report using additional priors and considerable engineering efforts. 

Fortunately, we can take advantage of explicitly provided material channels and avoid solving a difficult inverse rendering problem if we find a simpler way to pass $n$ channels to a 3-channel loss. 
We tested different more or less elaborated ideas such as computing a partial component analysis, training a $n$-to-3 channel encoder, etc.
In the end, we found that the best solution consists of choosing random triplets in the $n$ channels.
It provides a surprisingly simple and well-founded approach with stable outcome shown in Figure~\ref{fig:differential_rendering_loss}-(b).

\vspace{-2mm}

\clearpage
\section{Our Multi-Channel Textural Loss}
\label{sec:triplets}

\paragraph{The 3-channel loss.}

We build upon the 3-channel textural loss introduced by Gatys et al.~\shortcite{Gatys15}.
They define the textural distance between two RGB images $I_3$ and $\tilde I_3$ as the MSE between the Gram matrices of the activations produced by the images in the $L$ layers of a pretrained CNN:
\begin{align}
\label{eq:3_channel_loss}
\mathcal{L}_{3\text{-channel}}\left(I_3, \tilde I_3\right) = \sum_{l=1}^L \frac{1}{N_l^2}\| G^l - \tilde G^l\|^2,
\end{align}
where $G^l$ and $\tilde G^l$ are the Gram matrices of the $N_l$ deep features extracted from respectively $I$ and $\tilde I$ at layer $l$ in the pretrained CNN.
In our experiments, we use a pretrained VGG-19~\cite{Simonyan14}.

\paragraph{Combining multiple 3-channel losses.}

\begin{figure}[!h]
\begin{tikzpicture}
\draw (0,0) node[rectangle,draw,inner sep=0] {\includegraphics[width=0.225\linewidth]{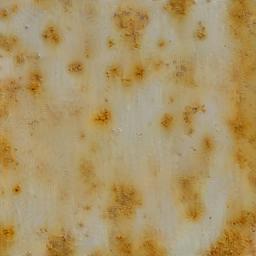}};
\draw (2,0) node[rectangle,draw,inner sep=0] {\includegraphics[width=0.225\linewidth]{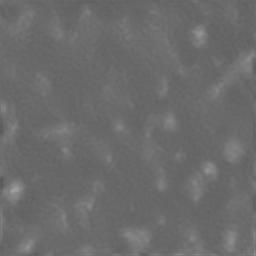}};
\draw (4+0.1,0) node[rectangle,draw,inner sep=0] {\includegraphics[width=0.225\linewidth]{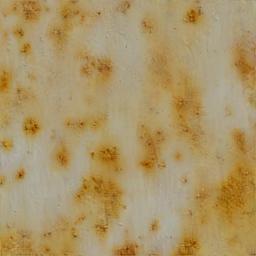}};
\draw (6+0.1,0) node[rectangle,draw,inner sep=0] {\includegraphics[width=0.225\linewidth]{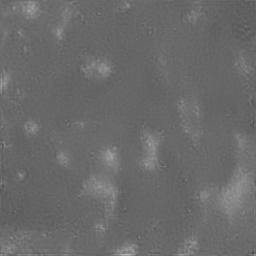}};
\draw[black] (0-0.05,-0.55) circle (5pt);
\draw[black] (2-0.05,-0.55) circle (5pt);
\draw [->] (0-0.05+0.1, -0.55) -- (2-0.05-0.1, -0.55);
\draw (1.7, -0.3) node {\scalebox{1}{\color{myred}\ding{55}}};
\draw[black] (4.1-0.2,-0.58) circle (7pt);
\draw[black] (6.1-0.2,-0.58) circle (7pt);
\draw [->] (4.1-0.2+0.16, -0.58) -- (6.1-0.2-0.16, -0.58);
\draw (5.6, -0.3) node {\scalebox{1}{\color{mygreen}\ding{51}}};
\draw (0,-1.15) node {\tiny \textit{albedo}};
\draw (2,-1.15) node {\tiny \textit{roughness}};
\draw (4+0.1,-1.15) node {\tiny \textit{albedo}};
\draw (6+0.1,-1.15) node {\tiny \textit{roughness}};
\draw (-0.3,+1.3) node {\small (a)};
\draw (4.4,+1.3) node {\small (b)};
\draw (1.0,+1.4) node {\tiny $\mathcal{L}_{3\text{-channel}}$ on \textit{albedo}};
\draw (0.96,+1.2) node {\tiny $+ \mathcal{L}_{3\text{-channel}}$ on \textit{roughness}};
\draw (5+0.1,+1.3) node {\tiny $\mathcal{L}_{n\text{-channel}}$};
\draw [-] (3.05, -1.2) -- (3.05, +1.2);
\end{tikzpicture}
\vspace{-4mm}
\caption{\label{fig:cross_correlation} 
(a) Optimizing multiple $3$-channel losses on separate triplets produces textures that look realistic independently but that are not correlated, i.e. the spatial features do not match.
(b) Our loss preserves inter-channel correlations since it optimizes for all the possible channel combinations. 
}
\vspace{-5mm}
\end{figure}
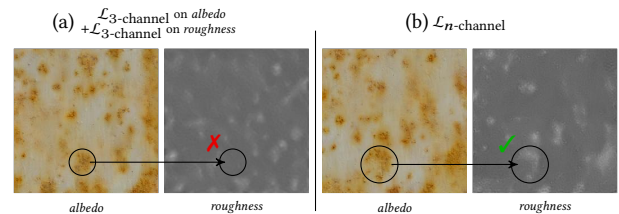

Accounting for more than 3 channels can be done by adding multiple 3-channel losses applied on the different maps. 
For instance, in Figure~\ref{fig:cross_correlation}-(a) we optimize for the sum of two 3-channel losses computed for the \emph{albedo} and the \emph{roughness} triplets separately.
This produces a texture whose \emph{albedo} and \emph{normal} look realistic separately but do not match together because the correlations between the albedo and the normal have not been accounted for. 

\paragraph{Our $n$-channel loss.}

In order to account for all the inter-channel correlations, we define the loss between two $n$-channel images $I_n$ and $\tilde I_n$ as the expectation of the 3-channel loss over \emph{all} the possible triplets:
\begin{align}
\label{eq:n_channel_loss}
\mathcal{L}_{n\text{-channel}}\left(I_n, \tilde I_n\right) = \mathbb{E}\left[\mathcal{L}_{3\text{-channel}}\left(\text{triplet}(I_n), \text{triplet}(\tilde I_n)\right)\right],
\end{align}
where $\text{triplet}\left(I_n\right)$ is a $3$-channel image whose channels are chosen randomly among the $n$ channels of $I_n$.
As shown in Figure~\ref{fig:cross_correlation}-(b), $\mathcal{L}_{n\text{-channel}}$ preserves inter-channel correlations.
The generated textures have the same feature at the same places channel-wide.
The downside is a direct evaluation of $\mathcal{L}_{n\text{-channel}}$ requires evaluating $\mathcal{L}_{3\text{-channel}}$ for all possible triplets and averaging the results. 
A material of $n$ channels hence requires $n^3$ different evaluations, which is untractable in practice. 

\paragraph{Stochastic evaluation.}

To overcome this problem we proceed as shown in Figure~\ref{fig:teaser}.
Instead of evaluating $\mathcal{L}_{3\text{-channel}}$ on all the possible triplets, we only evaluate it on a \emph{single} triplet that is randomized for each batch during learning.
In other words, we compute a stochastic estimate of Equation~(\ref{eq:n_channel_loss}):
\begin{align}
\label{eq:n_channel_loss_stochastic}
\hat{\mathcal{L}}_{n\text{-channel}}\left(I_n, \tilde I_n\right) = \mathcal{L}_{3\text{-channel}}\left(\text{triplet}(I_n), \text{triplet}(\tilde I_n)\right).
\end{align}
This evaluation is fast, practical and does not change the optimum of the $L^2$ minimization because the estimate is unbiased.
Furthermore, it remains robust because the randomness induced by the stochastic evaluation is similar to the natural randomness of stochastic gradient descent and well-handled by state-of-the-art optimizers.

\section{Training and results}

\begin{figure}[h]
\begin{tabular}{@{\hspace{-2mm}} c @{}}
\begin{tikzpicture}
\draw (-3.4-2*0.12,1.2) node {\small example 1};
\draw (0-2*0.12,1.2) node {\small generated};
\draw (+3.4-2*0.12,1.2) node {\small example 2};
\draw (+3.4-2*0.12,0+2*0.12) node[rectangle,draw,inner sep=0] {\includegraphics[height=0.17\linewidth]{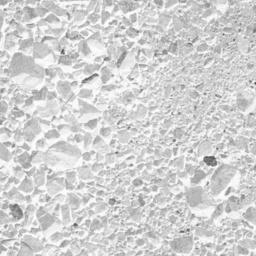}};
\draw (+3.4-1*0.12,0+1*0.12) node[rectangle,draw,inner sep=0] {\includegraphics[height=0.17\linewidth]{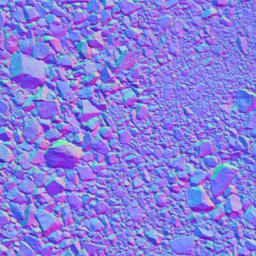}};
\draw (+3.4-0*0.12,0+0*0.12) node[rectangle,draw,inner sep=0] {\includegraphics[height=0.17\linewidth]{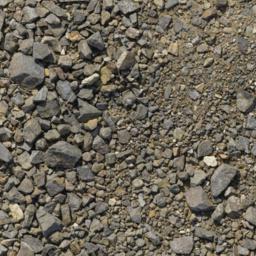}};
\draw (0-2*0.12,0+2*0.12) node[rectangle,draw,inner sep=0] {\includegraphics[height=0.17\linewidth]{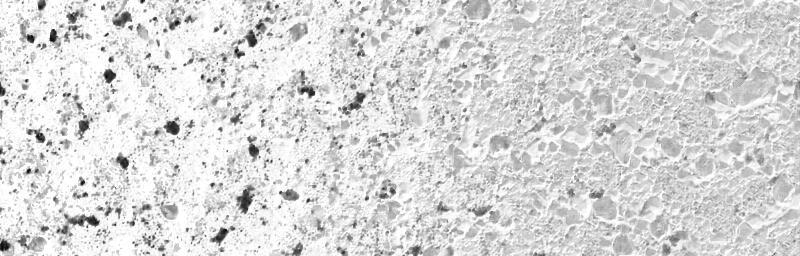}};
\draw (0-1*0.12,0+1*0.12) node[rectangle,draw,inner sep=0] {\includegraphics[height=0.17\linewidth]{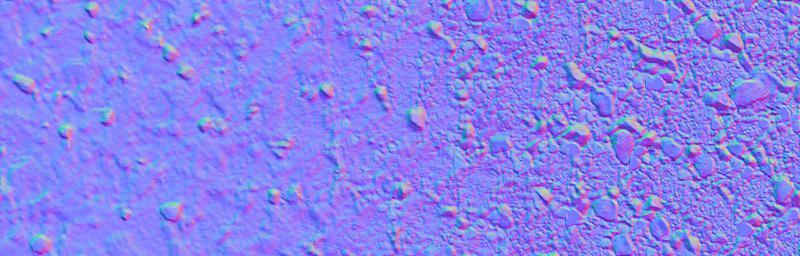}};
\draw (0-0*0.12,0+0*0.12) node[rectangle,draw,inner sep=0] {\includegraphics[height=0.17\linewidth]{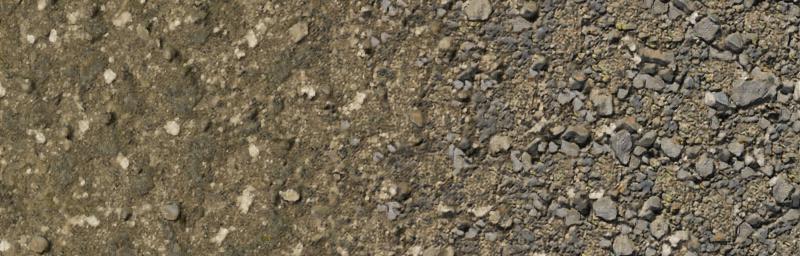}};
\draw (-3.4-2*0.12,0+2*0.12) node[rectangle,draw,inner sep=0] {\includegraphics[height=0.17\linewidth]{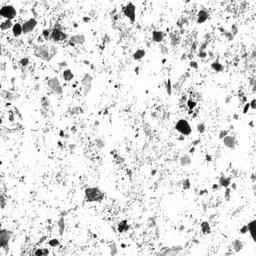}};
\draw (-3.4-1*0.12,0+1*0.12) node[rectangle,draw,inner sep=0] {\includegraphics[height=0.17\linewidth]{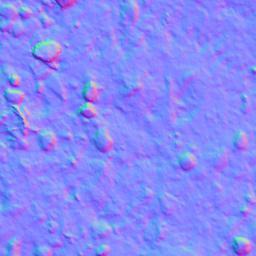}};
\draw (-3.4-0*0.12,0+0*0.12) node[rectangle,draw,inner sep=0] {\includegraphics[height=0.17\linewidth]{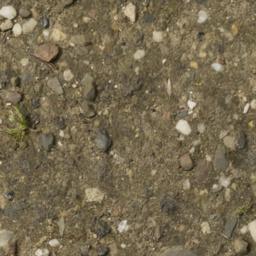}};
\end{tikzpicture}
\end{tabular}
\vspace{-5mm}
\caption{\label{fig:interpolation} 
We train a multi-texture generator that allows for interpolating material textures.
We trained the same architecture for 16 textures that includes these 2 examples.
}
\vspace{-3mm}
\end{figure}
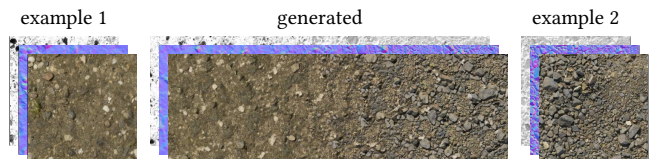

We use our loss as a drop-in extension of a 3-channel loss to train generative architectures of $n$ rather than $3$ channels. 
We train a mono-texture~\cite{UlyanovV1} and a multi-texture~\cite{Li2017} generators, which we adapted to output $n$ channels. 
We use $\hat{\mathcal{L}}_{n\text{-channel}}$ for sole loss function without further priors or regularization terms and we train with the Adam optimizer.
Note that we use a vanilla implementation of the 3-channel loss of Gatys et al.~\shortcite{Gatys15}.
Several improvements to this loss have been published and implementing them would directly benefit to our $n$-channel extension as well.  
Figure~\ref{fig:teaser} shows a result generated by our mono-texture generator.
Figure~\ref{fig:interpolation} shows a result generated by our multi-texture generator.
These generative architectures are capable of producing arbitrarily-large texture at inference time with variation (no verbatim copying of the exemplar).
Our supplemental material shows further results with different sets of channels on various textures.

\section{Conclusion}

We have proposed a simple approach to extend existing 3-channel textural losses to $n$ channels.  
The main idea is to span all the possible inter-channel correlations thanks to a stochastic evaluation that can be implemented in a single line of code.
With this approach we hope to bring a vast literature of neural texture synthesis approaches to material texture synthesis without further efforts.

\vspace{-1mm}

\bibliographystyle{ACM-Reference-Format}
\bibliography{siggraph}
\end{document}